\documentclass[final]{article}

\usepackage{arxiv}

\usepackage[utf8]{inputenc} 
\usepackage[T1]{fontenc}    
\usepackage{hyperref}       
\usepackage{url}            
\usepackage{booktabs}       
\usepackage{amsfonts}       
\usepackage{nicefrac}       
\usepackage{microtype}      
\usepackage{multicol}
\usepackage{graphicx}
\usepackage{comment}

\usepackage{natbib}
\setcitestyle{authoryear,aysep={,},round,citesep={;}}

\usepackage{authblk}

\usepackage{xcolor}
\usepackage{hycolor}
\hypersetup{colorlinks=true,citecolor=[rgb]{0,0.18,0.65}, 
urlcolor=[rgb]{0,0.18,0.65}, linkcolor=[rgb]{0,0.18,0.65}, final = true}
\usepackage{doi}

\def\GR{\mathrm{GR}}
\def\CD{\mathrm{CD}}
\def\SG{\mathrm{SG}}
\def\Pasym{P_\mathrm{asym}}
\def\Psig{P_\mathrm{sig}}
\def\Padj{P_\mathrm{adj}}
\def\Ber{\mathrm{Ber}}
\def\Csigi{C^{\mathrm{sig}}_{i}}
\def\Csigj{C^{\mathrm{sig}}_{j}}
\def\Tdiv{T^{\mathrm{div}}}
\def\Tsig{T^{\mathrm{sig}}}
\def\Erand{E_{\mathrm{rand}}}
\def\Rec{\mathrm{Rec}}
\def\Recall{\mathrm{Rec}^{\mathrm{all}}}
\def\Cfate{C^{\mathrm{fate}}}
\def\Creg{C^{\mathrm{reg}}}

\definecolor{YKB}{rgb}{0.00,0.18,0.65}

\title{Large-Scale Survey of Cell-Differentiation Programs 
in a Generative Model Reveals Regeneration 
as an Epiphenomenon of Development}

\author[ 1,*]{Somya Mani}
\author[ 1,*]{Tsvi Tlusty}
\affil[1]{Institute for Basic Science -- Center for Soft and Living Matter, Ulsan-44919, South Korea}
\affil[*]{Correspondence: somyamn@gmail.com (S.M.), tsvitlusty@gmail.com (T.T.)}

\begin{document}
\maketitle

\begin{abstract}
Development combines three basic processes --- asymmetric cell division, signaling and gene regulation --- in a multitude of ways to create an overwhelming diversity of multicellular life-forms. 
Here, we attempt to chart this diversity using a generative model. 
We sample millions of biologically feasible developmental schemes, allowing us to comment on the statistical properties of cell-differentiation trajectories they produce. Our results indicate that, in contrast to common views, 
cell-type lineage graphs are unlikely to be tree-like. 
Instead, they are more likely to be directed acyclic graphs, 
with multiple lineages converging on the same terminal cell-type. 
Additionally, in line with the hypothesis that whole body regeneration 
is an epiphenomenon of development, 
a majority of the ‘organisms’ generated by our model can regenerate using pluripotent cells. 
The generative framework is modular and flexible, 
and can be adapted to test additional hypotheses about general features of development.
\end{abstract}

\vspace{6mm}

\begin{multicols}{2}

\section{Introduction}
Contrary to intuition, the key molecules and mechanisms that go into the development of a human ( >200 cell-types \citep{milo2009bionumbers}) are the same as those required to produce a hydra (just 7 cell-types \citep{hwang2007evolutionary}). More generally, there is a huge diversity of forms and complexity across multicellular organisms, but key molecules of development in \textit{Metazoa} and in multicellular plants are conserved across the respective lineages \citep{meyerowitz2002plants}. The basis of this diversity is illustrated by mathematical models of development which explore possible mechanisms of producing distinctive patterns found in different organisms, for example, segments in \textit{Drosophila} \citep{von2000segment}, stripes in zebrafish \citep{volkening2015modelling}, and dorso-ventral patterning in \textit{Xenopus} larvae \citep{ben2014scaling}. At a much broader scale, single cell transcriptomics and lineage tracing techniques have made it possible to map the diversity of forms of extant multicellular organisms \citep{kester2018single}. Here, we ask about the limits of diversity that development can generate. And reciprocally, we ask what is common among all organisms that undergo development.

Biological development is modular \citep{bolker2000modularity}, and its outcome rests on gene regulation that is switch-like, rather than continuous \citep{albert2003topology,garfield2013impact}. Keeping this in mind, we constructed a generative model of development with three basic ingredients: asymmetric cell division, signaling and gene regulation \citep{alberts2002universal}. Although much is known about the detailed molecular machinery of development \citep{gilbert2017developmental}, naturally, these details come from studies on a few model organisms. We choose to not include all these important particular features in our model for the sake of efficiently and systematically sampling a broad space of developmental schemes. Nonetheless, our model is capable of expressing specific examples of known developmental pathways, which we demonstrate using the \textit{Drosophila} segment polarity network analysed in \cite{albert2003topology}.

We encode organisms in our model as lineage graphs, which show differentiation trajectories of the various cell-types in the organism. Traditionally, mathematical models in the literature elucidate developmental mechanisms responsible for known differentiation trajectories \citep{sharpe2017computer}. Here we take the inverse approach, and at a much broader scale; we sample across millions of biologically plausible developmental rules and map out the lineage graphs they produce. By tuning just three biologically meaningful parameters --- which control signaling, cellular connectivity and cell division asymmetry --- our model produces a rich collection of organisms with diverse cell-type lineage graphs, ranging from those with a single cell-type, to organisms with close to a hundred cell-types. Notably, tree-like lineage graphs are rare in our model. This could indicate that, contrary to popular belief, lineage graphs of real organisms are not tree-like; they are more likely to be directed acyclic graphs (DAGs). Additionally, an unanticipated outcome of our model is that most organisms we generate are capable of whole body regeneration. Our result supports the hypothesis that regeneration is an epiphenomenon of development \citep{goss1992evolution}. Despite the coarse-grained nature of our model, it generates 'organisms' that reproduce hallmarks of real biological organisms. The model also produces concrete predictions, and we discuss how these predictions can be experimentally tested on animals like \textit{Planaria}, in which regeneration is based on adult pluripotent cells \citep{reddien2018cellular}.

\section{Generative model of development}
\label{sec:headings}


\begin{figure*}[htbp!]
    \centering
    \includegraphics[scale=0.54]{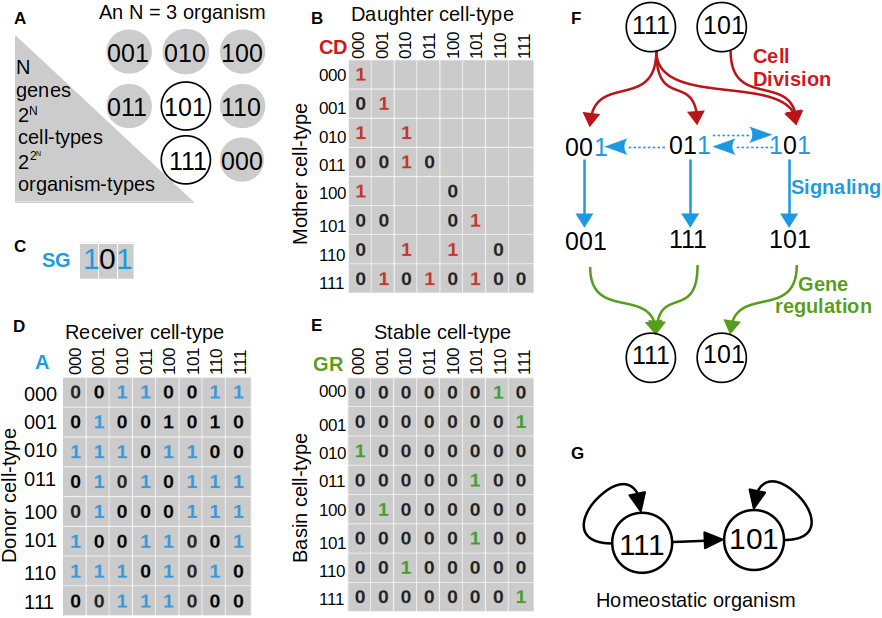}
    \caption{\textbf{Generative Model}. \textbf{(A)} An organism with N=3 genes and two cell-types. Circles represent all possible cell-types. The organism is composed of cell-types represented by white circles, and does not contain the grey cell-types. Binary strings written inside the circles represent the presence (1) or absence (0) of gene products in those cell-types. \textbf{B},\textbf{C},\textbf{D} and \textbf{E} describe the rules for development of the organism in \textbf{A}. \textbf{(B)} Cell division matrix $\CD$. For all $j$ such that $\CD(i,j) = 1$, cell-type $i$ produces cell-type $j$ upon cell-division. \textbf{(C)} Signaling matrix $\SG$. Genes 1 and 3, which are labelled in blue, produce signaling molecules. \textbf{(D)} Signaling adjacency matrix $A$. $A(i,j) = 1$ implies that cell-type $j$ receives all signals produced by cell-type $i$. \textbf{(E)} Gene regulation matrix $\GR$. $\GR(i,j)=1$ implies that cell-type $j$ is a stable cell-type, and cell-type $i$ maps to cell-type $j$. \textbf{(F)} Schematic of 'organismal development' in the model. All cell-types synchronously undergo cell-division according to $\CD$, the daughter cells exchange signals according to $\SG$ and $A$, and cells respond to signals through gene regulation according to $\GR$. The process repeats until it reaches a steady state. Here we show how the homeostatic organism in \textbf{A} is obtained using the developmental rules matrices in \textbf{B,C,D} and \textbf{E}. \textbf{(G)} Lineage graph of the homeostatic organism in \textbf{A}.
    }

    \label{fig:Fig1}
\end{figure*}

Organisms in the model contain genomes with $N$ distinct genes. By `Genes', we refer not to single genes, but to gene regulatory modules that control cellular differentiation \citep{mochizuki2013dynamics}. 
In different cell-types of an organism, products of different sets of genes can be present (1) or absent (0). We represent a cell-type as a $N$-length binary string. For example, for $N=3$, a cell-type $C=[101]$ contains products of genes 1 and 3 but not gene 2 products. 
(In Supplementary section 6.9, we demonstrate how 'Genes' can also be used to encode spatial information using the well-known \textit{Drosophila} segment polarity network as an example(Figs.\ref{fig:figs10}, \ref{fig:figs11}).

Cell-types are ordered according to standard binary ordering, i.e., the cell $[101]$ can equivalently be written as $C_{5}$. We only look at whether a given cell-type is present or absent in organisms, rather than the number of cells of any given cell-type. Therefore, since each of the $N$ genes can be either 1 or 0, there are at most $2^N$ distinct cell-types in an organism, and $2^{2^N}$ cell-type compositions for organisms (Fig.\ref{fig:Fig1}(A)). Note that the number of distinct organisms is larger than $2^{2^N}$, since different organisms may have the same set of cell-types but distinct lineage graphs (Fig.\ref{fig:Fig1}(G)).) 

We represent development as a repeated sequence of cell division, intercellular signaling, and gene regulation: 

\textit{Cell division.---} Cells in the model undergo asymmetric cell-division. Although in real multicellular organisms, a single cell only divides into two daughter cells, a single cell-type may represent a population of cells, which need not all behave in the same way \citep{altschuler2010cellular, klein2011universal}. We capture this heterogeneity by allowing cells in our model to divide into more than two types of daughter cells. Asymmetry of cell division is controlled by the parameter $\Pasym \in [0,1]$, which is the probability that a daughter cell does not inherit the product of a given gene from the mother cell. That is, $\Pasym = 0$ implies symmetric division, where all daughter cells inherit all gene products from the mother cell, and at $\Pasym = 1$, no daughter cell receives any gene products from the mother cell.  We assume that in the instant directly following division, no new gene products are formed. Therefore, genes that were in a 1 state in the mother cell can switch to a 0 state in daughter cells due to unequal and insufficient partitioning of the gene product during division \citep{knoblich2008mechanisms}, but genes that are in a 0 state in the mother cell are necessarily in a 0 state in the daughter cells as well. For any given organism in the model, we predetermine the sets of daughter cells produced by different cell-types randomly according to $\Pasym$, and encode this in a binary matrix $\CD$ (Fig.\ref{fig:Fig1}(B)).

\textit{Signaling.---} The number of distinct signaling molecules in an organism is controlled by the parameter $\Psig \in [0,1]$, which is the probability that the product of any particular gene is a signaling molecule. Parameter $\Padj \in [0,1]$ controls signal reception; a cell-type $C_{i}$ can receive signals from a cell-type $C_{j}$ with probability $\Padj$. As in the case of cell-division, for each organism, the set of signaling molecules, and the pairs of cells that are allowed to exchange signals are predetermined and stored in a binary vector $\SG$, and a binary matrix $A$, respectively (Fig.\ref{fig:Fig1}(C,D)). Cells can only receive signals from other cell-types present in the same time step, and recipient cells receive all the signals produced by donor cells. In recipient cell-types, in response to incoming signals, the corresponding genes are set to a 1 state (Fig.\ref{fig:Fig1}(F)).

\textit{Gene regulation.---} We model gene regulation as random Boolean networks (RBNs) \citep{gershenson2004introduction}; the states of genes depend on each other through arbitrarily complex Boolean rules. Updates in gene states result in updates in cell-types. In this scheme, some cell-types update to themselves (stable states), and other cell-types ultimately update to one of the stable cell-types, that is, they lie in the basin of a stable state. Here, instead of encoding RBNs explicitly, we describe gene regulation directly as the set of stable cell-types and their basins. For each organism, we predetermine its gene regulation and encode it in a binary matrix $\GR$ (Fig.\ref{fig:Fig1}(E)). Our model is deterministic; once the matrices $\CD$, $\SG$, $A$ and $\GR$ are determined for an organism, they remain fixed for the rest of the simulation. The model is also synchronous; all cell-types in the organism divide simultaneously, after which the developing organism is composed only of all daughter cells produced in this step. These daughter cells simultaneously exchange signals, in response to which the states of all the genes, in each daughter cell-type are updated simultaneously according to $\GR$ (Fig.\ref{fig:Fig1}(F)). A time-step in the model represents a single repeat of cell-division, signaling and gene regulation.

The process of development ends when the set of cell-types in a developing organism repeats itself. We call this set of cell-types the steady state of the organism, and the number of time-steps between two repeats the period of the steady state. Since this is a finite and deterministic system, starting from any initial condition, such a steady state can always be reached. We call period-1 steady states \emph{homeostatic organisms} (Fig.\ref{fig:Fig1}(F)). Although organisms with complex, period>1 life-cycles, such as land plants with alternation of generation \citep{graham2000origin} exist in nature, in this study, we focus on homeostatic organisms.
We represent homeostatic organisms as their \emph{cell-type lineage graphs} (Fig.\ref{fig:Fig1}(G)). The nodes of this graph represent cell-types in the homeostatic organism, and directed edges represent lineage relationships between these cell-types. Let some cell-types $A$ and $B$ in a homeostatic organism be represented by nodes $V_{a}$ and $V_{b}$, respectively, in its lineage graph. Then, there is an edge from $V_{a}$ to $V_{b}$ if one of the daughter cells of $A$ gives rise to $B$ after one round of cell-signaling and gene regulation. Note that the lineage graphs in the model are for the adult homeostatic organism, and do not represent developmental trajectories which map transitions of embryonic cell-types.

\section{Results}

\subsection{Homeostatic organisms span a large range of sizes} 
\begin{figure*}[htbp!]
    \centering
    \includegraphics[scale=0.58]{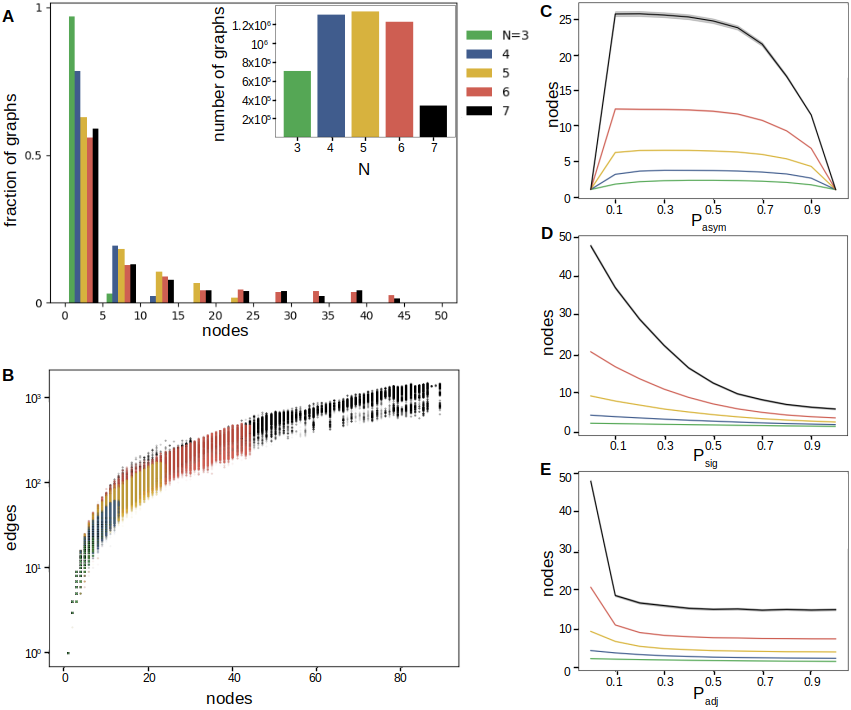}
    \caption{\textbf{Diversity of lineage graphs.}
   \textbf{(A)} Histogram of number of nodes in lineage graphs obtained with different N. Histogram bins are of size 5. Inset: number of lineage graphs in the data at different values of N. 
   \textbf{(B)} scatter plot of number of edges and number of nodes in lineage graphs. Transparency has been added to points to make density of points more apparent. (\textbf{C}, \textbf{D} and \textbf{E}) Number of nodes in lineage graphs obtained at different $N$ as a function of \textbf{(C)} $\Pasym$, \textbf{(D)}  $\Psig$ and \textbf{(E)} $\Padj$ (see also Fig.\ref{fig:supp_fig1}). Thick lines represent the mean and shaded regions around the lines represent standard deviation (the shaded regions are hard to see because the standard deviations are low). 
   }
    \label{fig:Fig2}
\end{figure*}

We looked at millions of homeostatic organisms, spanning systems with $N = [3,4,5,6,7]$ number of genes (Fig.\ref{fig:Fig2}(A),inset). Therefore, the largest possible organism in our data can contain $2^7 = 128$ cell-types. 99.88\% of these homeostatic organisms had lineage graphs with a single connected component. In the following, we describe lineage graphs of these single-component homeostatic organisms. While a majority of graphs in our data are small (1-5 nodes), the largest graphs have 89 nodes (Fig.\ref{fig:Fig2}(A)). Naturally, the number of edges in lineage graphs increases with the number of nodes, but this increase is slower than that expected for simple random graphs (Fig.\ref{fig:Fig2}(B), Fig.\ref{fig:supp_fig1}(A)). 
The number of nodes in lineage graphs follows closely the diversity of daughter cell-types produced (Fig.\ref{fig:supp_fig1}(C,D)). At very low $\Pasym$, cells produce daughters cells identical to themselves, and at very high $\Pasym$, most daughter cells are of the type $[0,0,...0]$. Therefore at these values, diversity of daughter cells, and correspondingly the number of nodes in lineage graphs, is low. At other values of $\Pasym$, the number of nodes stays level and decreases slowly beyond $\Pasym = 0.5$ (Fig.\ref{fig:Fig2}(C)). Number of nodes decreases as $\Psig$ increases (Fig.\ref{fig:Fig2}(D)). Intuitively, high levels of signaling causes genes in a 1 state to `spread out', effectively leading to a homogenization of cell-types. The sharp decrease in the number of nodes in response to increase in $\Padj$ indicates that a low level of inter-cellular connectivity is sufficient for signals to percolate throughout the organism (Fig.\ref{fig:Fig2}(E), Fig.\ref{fig:supp_fig1}(B)).


\subsection{Diversity of lineage graph topologies and the dearth of tree-like lineage graphs}
Paths in a lineage graph represent differentiation trajectories of the organism's cell-types. Here, we classify lineage graphs into  six topologies, each of which contain qualitatively different paths: (i) unicellular (single cell-type), (ii) SCC (Strongly Connected Component -- all paths are cyclic), (iii) cyclic (contains both cyclic and acyclic paths), (iv) chains (acyclic graphs with no branches), (v) trees (acyclic graphs with branches) and (vi) DAGs (Directed Acyclic Graphs, which contain edges connecting different branches. These edges represent the convergence of multiple cell-lineages to the same terminal cell-type). We ignore self-edges during lineage graph classification. 

\begin{figure*}[htbp!]
    \centering    \includegraphics[scale=0.445]{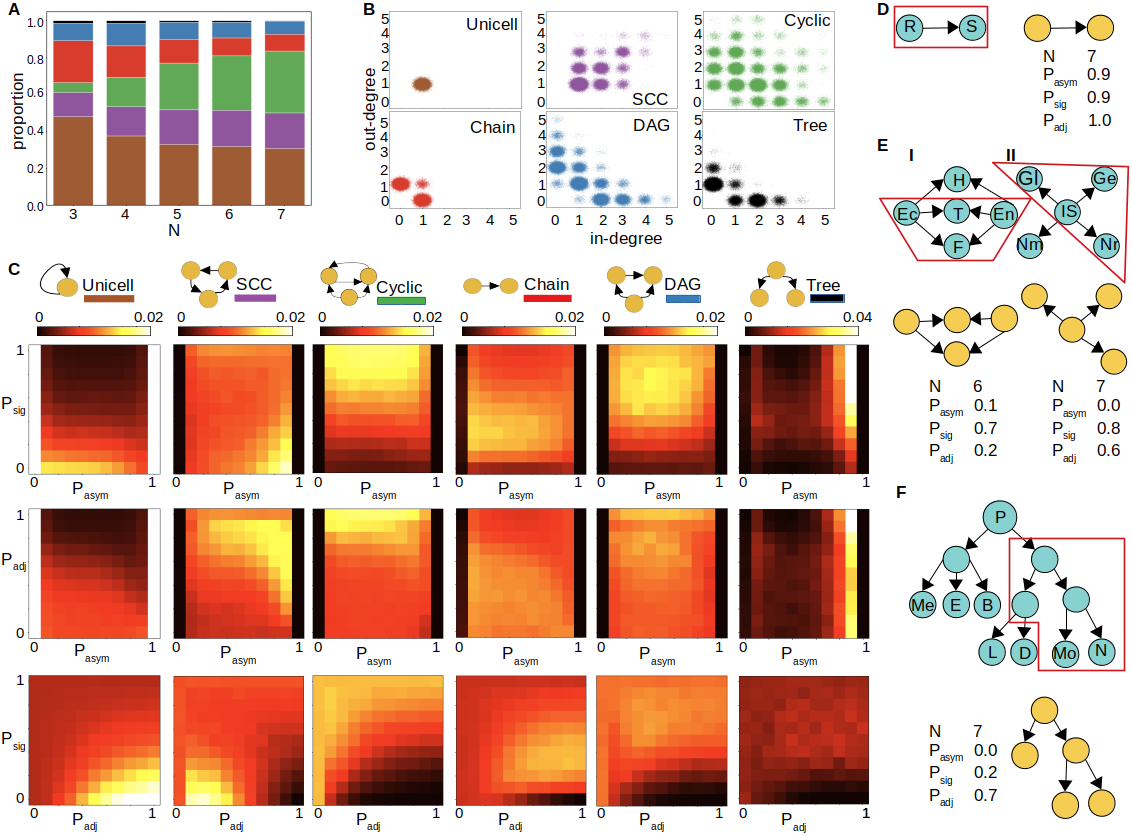}
    \caption{\textbf{Lineage graph topologies} 
    \textbf{(A)} Stacked histogram for topologies of lineage graphs obtained with different N (see also Fig.\ref{fig:figs2}). Different topologies are represented with different colours: unicellular:brown, SCC:purple, cyclic:green, chain:red, DAG:blue, tree:black. Heights of colored blocks represent the proportions of corresponding topologies. 
    \textbf{(B)} Scatter plots for in-degrees and out-degrees of graph nodes in different topologies. Noise has been added to points in the plots to make the density of points at each position more apparent. 
    \textbf{(C)} 2-D histograms indicating distribution of topologies across parameter space. The first row of histograms show distributions along $\Pasym$ and $\Psig$, second row along $\Pasym$ and $\Padj$, and the third row along $\Padj$ and $\Psig$. Different columns correspond to histograms for different topologies, as indicated at the top of each column. Intensity of colours in histograms in any column indicates the fraction of graphs of a particular topology found in the corresponding parameter region, according to the colourbars given at the top of each column. \textbf{(D,E,F)} Examples of lineage graphs of real organisms. Circles represent cell-types, and edges represent lineage relationships between cell-types. Graphs with blue circles belong to real organisms, and graphs with yellow circles are model generated lineage graphs that are of the same graph-type (chain, DAG, tree, etc.), and best resemble the corresponding real lineage graphs. Parameter values ($N,P_{asym}, P_{sig},P_{adj}$) where these yellow graphs can be found are indicated in the figure. Parts of real lineage graphs that perfectly match the model's lineage graphs are shown in red boxes. \textbf{(D)} \textit{Volvox} has a chain-type lineage graph. Key to cell-types: R: reproductive cells, S: somatic cells \citep{matt2016volvox}. \textbf{(E)} \textit{Hydra}. Its lineage graph has two components; I and II. I is a DAG-type graph, and II is a tree-type graph. We treat these two components as separate graphs. Key to cell-types: Ec: ectodermal epithelial stem cell, En: endodermal epithelial stem cell, IS: interstitial stem cell, H: hypostome, T: tentacle, F: foot, Gl: gland cells, Ge: germ cells, Nm: nematocyst, Nr: neuron \citep{siebert2019stem}. \textbf{(F)} human hematopoeitic system has a tree-type lineage graph. Key to cell-types: P: progenitor cells, Me: megakaryocytes, E: erythrocytes, B: basophils, L: lymphocytes, D: dendritic cells, Mo: monocytes, N: neutrophils \citep{pellin2019comprehensive}. 
     }
    \label{fig:Fig3}
\end{figure*}

In our data, unicellular graphs are the most abundant (36\%). Acyclic graphs (chains, trees and DAGs) comprise about 25\% of our graphs. Of these, trees are the rarest (<1\% across all graphs) and chains are the most abundant (14.3\% across all graphs) (Fig.\ref{fig:Fig3}(A)). Although all topologies are spread widely across parameter space, different topologies are enriched in different regions of parameter space (Fig.\ref{fig:Fig3}(C)). No parameter region is monopolized by a single topology, except at extreme values of $\Pasym$, where, as discussed earlier, most graphs are unicellular. To a large extent, these topologies can be characterized by their in-degree and out-degree distributions. For instance, in chains, in-degrees and out-degrees are at most 1, whereas in SCCs, in-degrees and out-degrees are at least 1 (Fig.\ref{fig:Fig3}(B)). Therefore, for the most part, we can explain the model's propensity to generate certain topologies, in terms of its propensity to generate certain in-degree and out-degree distributions. However, we find that acyclic graphs are slightly more enriched in our data than in randomized graphs with the same in-degree and out-degree distributions (Fig.\ref{fig:figs2}).

To test whether graphs produced by our model are realistic, we compared our lineage graphs with those of real organisms (Fig.\ref{fig:Fig3}(D,E,F)). 98\% of all chain-type graphs in our data match exactly the \textit{Volvox} lineage graph (Fig.\ref{fig:Fig3}(D)). While we do not find model generated graphs which exactly match the lineage graphs for \textit{Hydra} and the human hematopoeitic system, we do find graphs that are identical to parts of the real lineage graphs (Fig.\ref{fig:Fig3}(E,F)). Recently, Plass \textit{et al.} performed lineage reconstruction for the whole adult planarian worm, and report the best supported spanning tree for its lineage graph \citep{plass2018cell}. We are unable to include this graph here, because the small sizes of our tree-like graphs makes a comparison unsuitable.

\subsection{Homeostatic organisms contain pluripotent cells}
We wanted to test whether these `homeostatic organisms' could self-reproduce. As a test, we looked at whether single cell-types taken from homeostatic organisms develop into the same organism using the same rules ($\GR$, $\CD$, $A$ and $\SG$) used to generate the organism from a random initial cell-type. In other words, we looked for pluripotent cell-types. We find that in about 97\% of all homeostatic organisms (and 95.2\% non-unicellular organisms) there is at least one such pluripotent cell-type. This is surprising, since cells, taken out of the context of signaling from other cell-types in the organism, are not expected to be regenerative. Additionally, regeneration trajectories, starting from these pluripotent cell-types tend to be short (Fig.\ref{fig:figs7}).

We tested whether this high level of pluripotency could be a trivial consequence, arising because perhaps the lineage graphs sampled in our data are the most probable graphs produced by these dynamics. But we find that in about 73.3\% of lineage graphs (73.1\% non-unicellular graphs), cell-types that are taken from homeostatic organisms are much more likely to generate it than cell-types not present in the homeostatic organism (Fig.\ref{fig:Fig4}(A)). 
We therefore measure \textit{regenerative capacity} of an organism as the fraction of pluripotent cells in the organism divided by the fraction of all cell-types (irrespective of whether it is a part of the organism, or not) that generate the organism. We call an organism \textit{regenerative} if its regenerative capacity is greater than 1.

Regenerative capacity differs among different topologies (Fig.\ref{fig:Fig4}(B), Fig.\ref{fig:figs6}). In particular, most tree-type graphs have low regenerative capacity. Regenerative capacity also depends on model parameters: at $\Pasym =0$, where cells divide to produce identical daughter cells, as expected, organisms are maximally regenerative. At $\Pasym =1$, where all cell-types produce the same daughter cell-type ($[0,0,...0]$), regenerative capacity is lowest (Fig.\ref{fig:Fig4}(C)). Regenerative capacity increases with $\Psig$ and $\Padj$ (Fig.\ref{fig:Fig4}(D,E)).

\subsection{Pluripotent cells removed from their organisms retain their cell fates}

In order to find the source of the high regenerative capacity of organisms in our data, we test how much the fate of a cell in the model depends on signaling. We define the fate of a cell-type $C$ in a homeostatic organism as the set of all cell-types that receive an edge from cell-type $C$ in the lineage graph of the organism. We call a cell-type \emph{independent} if it has the same cell-fate when taken out of the homeostatic organism, as it does within the organism. Note that \textit{pluripotency} and \textit{independence}, while related, are not synonymous; the differentiation trajectory of a pluripotent cell during regeneration could be different than its differentiation trajectory during homeostasis.

\begin{figure*}
    \centering
    \includegraphics[scale=0.45]{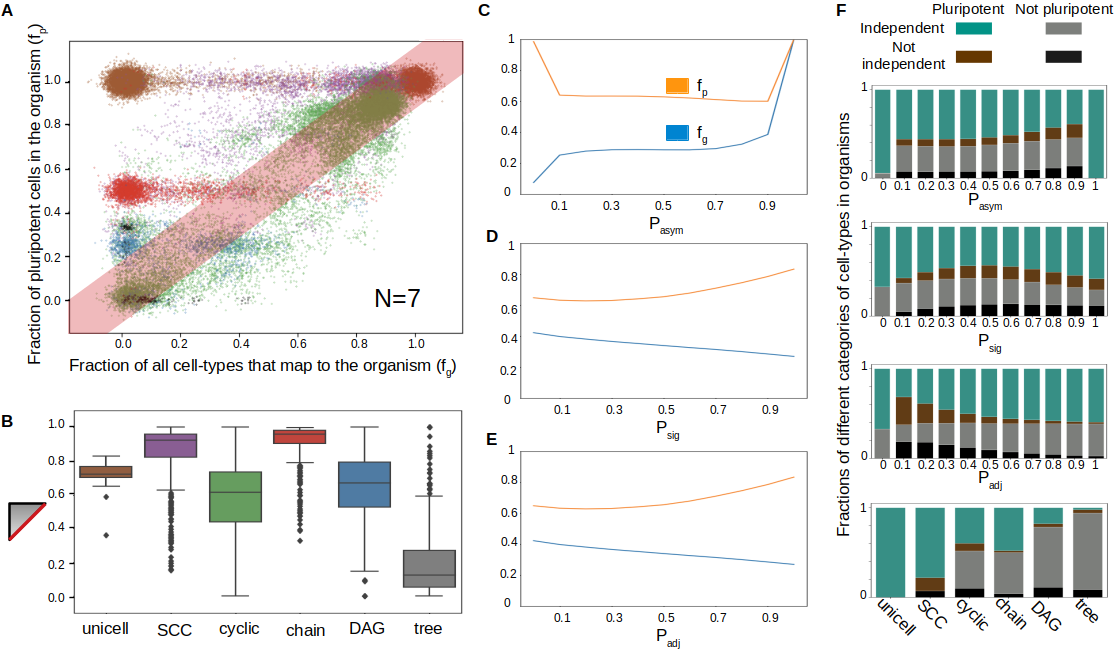}
    \caption{\textbf{Regenerative capacity.}
    \textbf{(A)} Scatter plot showing regenerative capacity for all $N=7$ organisms generated with a fixed gene-regulation matrix $\GR$ using different matrices $\CD$, $A$ and $\SG$ (for cell-division, cellular adjacency and signal transduction, respectively). Each point represents an organism. The x-axis is the fraction of all cell-types, those found in the organism, as well as those that are not, which develop into this homeostatic organism ($f_g$). The y-axis is the fraction of cell-types taken from within the organism which develop into this organism, i.e., the fraction of \textit{pluripotent} cells ($f_p$). Noise has been added to the position of points to make their density more apparent. Colours of points indicate the topology of their lineage graphs (as in Fig.\ref{fig:Fig3}): unicellular:brown, SCC:purple, cyclic:green, chain:red, DAG:blue, tree:black. Points above the red band are regenerative organisms (with $\frac{f_p}{f_g} > 1$).
    \textbf{(B)} Box plot of proportion of regenerative graphs of different topologies across all organisms in the data (see also Fig.\ref{fig:figs6}). For each $\GR$ used in our data, for a given graph topology, we looked at the fraction of graphs with regenerative capacity > 1 (equivalent to the fraction of points of a certain colour that occur above the red band in \textbf{A}). Boxes represent quartiles of the data set. Lines inside the box show the median, while whiskers show the rest of the distribution. Outliers are shown as diamonds. 
    \textbf{(C,D,E)} Variation of regenerative capacity across model parameters: \textbf{(C)} $\Pasym$, \textbf{(D)} $\Psig$, \textbf{(E)} $\Padj$. Fraction of pluripotent cells is shown in orange, and the fraction of all cells (present or absent from organisms) that develop into the organism is shown in blue. Bold lines represent mean values (shaded regions around the lines represent standard deviations, which are small and hardly noticeable). The average regenerative capacities of graphs at different parameter values can be judged by the difference in the heights between the orange and blue curves. \textbf{(F)} Stacked histograms for cell-types of different categories pooled from organisms across different parameter values (top 3), or across lineage graphs with different topologies (see also Fig.\ref{fig:figs8}, Fig.\ref{fig:figs9}). Different cell-type categories are represented with different colours. Non-pluripotent cells are represented in greys; independent: light grey, not independent: black. Pluripotent cells are represented in colours; independent: teal, not independent: brown. Heights of colored blocks represent the proportions of corresponding cell-types.
    }
    \label{fig:Fig4}
\end{figure*}

Surprisingly, we find that across all parameter regions, homeostatic organisms are enriched in independent cell-types (Fig.\ref{fig:Fig4}(F),Fig.\ref{fig:figs9}). About 62\% of all independent cells, pooled from all non-unicellular graphs, are pluripotent. That is, overall, independent cells are slightly more likely to be pluripotent than not. But, 85.2\% of all pluripotent cells, pooled from all non-unicellular graphs, are independent (see Fig.\ref{fig:figs8} for a breakdown according to number of pluripotent cell-types in an organism). This explains the most likely mechanism of pluripotency in our model: if cells produced by a single cell plucked out of the homeostatic organism also belong to the organism, the whole organism can be built up step-by-step starting from such a single cell.

Interestingly, while the proportion of independent cell-types pooled from all non-unicellular graphs is similar ($\geq 75\%$) across all topologies, different topologies have very different proportions of pluripotent cells (Fig.\ref{fig:Fig4}(F), fourth panel). Notably, in SCC-type lineage graphs, where all differentiation paths are cyclic, 99.8\% of all independent cells are pluripotent. Whereas in lineage graphs that contain acyclic differentiation paths, the proportion of pluripotent independent cells is lower; particularly in tree-type lineage graphs, where only 2.4\% of the independent cells are pluripotent. More generally, this indicates that not only the number of independent cell-types, but also their connectivity in the lineage graph is an important factor contributing to an organism's regenerative capacity. \\


\section{Discussion}
The process of development and its molecular mechanism is inherent in all \textit{metazoans} and in all plants \citep{meyerowitz2002plants}. This makes it difficult to design experiments to distinguish between emergent traits associated with development and traits that have evolved on top of it. Here, we have developed a minimal model where we can look at development in the absence of complications due to cross-talk with other biological processes. In our model, we include only those ingredients of development that are shared across all multicellular organisms, while not ascribing any particular form or mechanism to these processes. This allows us to identify traits that stem from the fact that the organisms undergo development, regardless of the details of the process. Note that such basic traits can still be subject to selection through regulatory processes on top of the key ingredients of development. We see such an emergent trait in our model: ability of \emph{whole body regeneration} (WBR) through pluripotency. WBR, though widely spread across basal metazoan phyla, is curiously absent in mammals and birds (\textit{Ecdysozoa}). Below, we discuss major assumptions and limitations of our model, and contrast these with mechanisms that occur in biological organisms which could effect regenerative ability.


\begin{itemize}

    \item \emph{Independent processes: }Cell-division, signaling, and gene regulation are treated as independent processes in the model. This is likely to be false in real animals. Primarily, this implies that not all regions of parameter space explored in this work are biologically feasible. In particular, cells in the model follow a simple program for asymmetric cell division that is intrinsic to cell-types, but extrinsic control of asymmetric cell division, involving cues from surrounding cells, does occur in animals~\citep{knoblich2008mechanisms}. Extrinsic control of asymmetric cell-division could lead to a decrease in the independence of cell-fates on cellular context which we see in the model, and thus lower regenerative capacity.     
    
    \item \emph{Chemical signaling:} We have encoded a form of signaling which depends on spatial ordering of cells: cell-types that are arranged \textit{closer} in some sense to other cell-types are adjacent to them and accept all the signals produced by them. Whereas, in real organisms, cells contain receptors that recognize signal molecules, rather than recognizing the donor cells that produce those signals. Firstly, since there are fewer kinds of signal molecules than there are cell-types, with this chemical recognition based signaling, on average, more cells are expected to exchange signals. In our model, level of signal exchange is controlled by $\Padj$, and we find that pluripotency increases with $\Padj$ (Fig.\ref{fig:Fig4}(E)). Therefore, it is likely that a switch to a chemical recognition based signaling will preserve the high level of pluripotency. Secondly, in the chemical recognition scheme, it is likely that a cell-type will receive the same set of signals even if some other cell-types in the organism are changed. That is, cell-fate is also likely to be more robust to changes in cellular context. Therefore, cell-types are also likely to be independent of cellular context as in the current model.
    
    \item \emph{Other schemes: }In the current model, we use the following scheme of development: cell-division, followed by signaling among daughter cells and gene regulation in response to signals exchanged. But there are other reasonable schemes which can also be considered. For example, a scheme where cell-division is followed by an additional step of gene regulation before signal exchange is also plausible. In the current model, daughter cells contain subsets of the contents of the  mother cell, and in this sense are more similar to each other than to daughters of other mother-cells. Therefore, in the current scheme, signals received from a sister cell are likely to be less \textit{effective} in changing cell-state than are signals received from other daughter cells. Gene regulation right after cell division would lead to a diversification of daughter cells, which is therefore likely to increase the level of effective signaling among daughter cells. But, as discussed earlier, we expect that an increase in the level of signal exchange to still lead to high regenerative capacity.
    
    \item \emph{Additional parameters:} The effect of processes such as asynchronous gene state updates \citep{chaves2005robustness}, and time delays involved in transfer of information about gene state updates \citep{cheng2013autonomous} has been tested on the \textit{Drosophila} segement polarity network, and found to have interesting effects on the robustness of phenotypes. Such processes could add to the richness of lineage graphs we obtain from our model, but come with the cost of additional parameters, which would limit the breadth of the sampling.
    
    \item \emph{Cell death:} Cells in the model do not die. Not including cell death in the model results in lineage graphs where each node has at least one out-edge.  We anticipate that including cell death would reduce the number of cycles in lineage graphs, leading to an increase in the proportion of acyclic graphs. Since regenerative capacity is linked to lineage graph topology, cell-death could be an important factor in determining regenerative capacity.

\end{itemize}

Although here we only provide intuitive arguments for what alternate versions of the model might yield, the framework of the model is easily amenable to manipulations, and differently constructed versions can be tested in the future.

The present model makes several predictions regarding general features of development and multicellular organisms. It suggests that presence of adult pluripotent cells should be a widespread trait in multicellular life-forms. In plants, we are already aware of pluripotent cells in the root and shoot meristems. But among animals, a wider investigation of regeneration and its mechanisms will be required to test this idea. A recent example of such a study is \citep{zattara2019phylum}, where the authors test the ancestral nature of regeneration in \textit{Nemertaean} worms, which are not classical model organisms. 

The distribution of lineage graph topologies in our data reflect the complexity and diversity of forms of multicellular animals that biological development is expected to produce. Under normal circumstances, cellular differentiation is expected to be irreversible. Therefore, we restrict our discussion here to the acyclic graphs that our model produces. Small (2-node) \emph{chains} are the most abundant acyclic lineage graphs in our data (Fig.\ref{fig:Fig3}(A), Fig.\ref{fig:figs3}(C)). In line with this, the simplest multicellular organisms, such as \textit{Volvox carteri} \citep{matt2016volvox}, an alga which evolved multicellularity only recently, has a chain like lineage graph. Interestingly, some cyanobacteria, such as \textit{Anabaena spaerica} \citep{claessen2014bacterial}, which display multicellularity during nitrogen starvation, also have chain-like lineage graphs.

Tree-type lineage graphs are rare in our data and tend to be small, and convergent rather than divergent (Figs. \ref{fig:Fig3}(A), \ref{fig:figs3}(E), \ref{fig:figs4}). This could indicate one of two things: This could imply that lineage graphs of complex organisms are unlikely to be tree-like. Our data suggests that they are more likely to be DAGs (directed acyclic graphs), i.e., organisms have higher levels of trans-differentiation than expected (Figs. \ref{fig:figs3}(D), \ref{fig:figs5}). Or, it could mean that more complex regulation, on top of the ingredients of this model, are at play in real organisms which lead to complex tree-like lineage graphs. A perhaps presumptuous, but interesting possibility is that tree-like lineage graphs were selected for because of their low regenerative capacity. There exist arguments and speculation over whether the \textit{Ecdysozoans} selectively lost the ability to regenerate, and why \citep{bely2010evolutionary}.

These questions surrounding the topologies of lineage graphs are likely to be resolved very soon in the future, given the rapid developments in single cell transcriptomics technology. A notable recent study is that of Plass \textit{et al.} \citep{plass2018cell}, where they assemble the whole organism lineage graph for \textit{Planaria}. A possible hurdle comes from the fact that current methods for lineage reconstruction using single cell transcriptomics data are not unbiased; in \citep{plass2018cell}, although lineage reconstruction yielded a complex graph, the authors highlight the best supported spanning tree of this graph. Current lineage reconstruction methods work best if a particular topology for lineage graphs is already anticipated, and most methods are designed to only find chains and trees \citep{saelens2019comparison,tritschler2019concepts}. A study by Wagner \textit{et al.} \citep{wagner2018single}, where single cell transcriptomics is used in conjunction with cellular barcoding, provides an example of a lineage reconstruction method which is unbiased towards particular topologies. In agreement with our result, the authors of this study found that zebrafish development is best represented by a DAG.

Lastly, we discuss how certain predictions of our work can be experimentally tested. Our results suggest that in organisms, such as \textit{Planaria}, where regeneration is based on adult pluripotent cells called c-neoblasts, these cells are likely to be independent of cellular context, that is their cell fates should not change when taken out of the body, or transplanted to other cellular contexts. c-Neoblast independence could explain the coarse pattern of distribution of specialized neoblasts across the planarian body, and also why the distribution of specialized neoblasts produced does not depend on which organ is amputated \citep{reddien2018cellular}. Recent development of a method to culture neoblasts in the lab \citep{lei2019cultured}, make it possible to experimentally test neoblast independence. Additionally, in the model, not only pluripotent cells, but also non-regenerative cells display independence. Therefore, we also predict that, at least in organisms such as \textit{Planaria}, cell-fate trajectories in organisms in homeostasis should reflect cell-fate trajectories in regenerating organisms. This can be addressed by lineage reconstruction experiments that compare lineage graphs of planarians in homeostasis with lineage graphs of regenerating planarians.

\section{Methods}

\subsection{Surveying the combinatorial space of developmental schemes}
We considered organisms with $N = \{3,4,5,6,7\}$ genes. For each $N$, we have looked at $\{ 100,100,100,92,25 \}$ randomly generated gene regulation matrices ($\GR$), respectively. For each $\GR$, all values from $[0,0.1,0.2,\dots,1.0]$ were used for the parameters $\Pasym$, $\Psig$ and $\Padj$. At each parameter value, 10 randomly chosen cell-types were used as initial conditions (in case of $N=3$, all 8 cell-types were used). A distinct set of developmental rules matrices ($\CD$, $A$ and $\SG$) was used in combination with each initial cell-type.  In all, we have looked at about $((100+100+100+92+25)\times11^{3}\times10) \approx 5.5\times10^6$ systems. 4858643 of these converged within 1000 time-steps into homeostatic organisms. 

\subsection{Model details}
\subsubsection{Asymmetric cell division}

In our model, for any cell-type $C_{i}$, we generate different sets of daughter cell-types $D_{i}$ using the parameter $\Pasym \in [0,1]$; for any daughter cell-type $D_{i_{i1}}\in D_{i}$, $\forall k \leq N$,
\begin{center}
 if $(C_{i}(k) = 0)$ then $(D_{i_{i1}}(k) = 0)$, and
 
 if $(C_{i}(k) = 1)$ then $(D_{i_{i1}}(k) = \Ber(\Pasym))$
\end{center}

We encode cell-division in a binary matrix $\CD_{2^N \times 2^N}$; $\CD(i,j) = 1$ if cell-type $C_{j} \in D_i$, else $\CD(i,j) = 0$ (Fig.\ref{fig:Fig1}(B)).  

\subsubsection{Signaling}
The probability that a gene in the model produces a signaling molecule is $\Psig\in [0,1]$. Formally, let $\SG = \{0,1\}^{N}$ be a binary vector. Then gene $k$ produces a signaling molecule if $\SG(k) =1$, where $\SG(k) = \Ber(\Psig)$ (Fig.\ref{fig:Fig1}(C)). Let $\SG_{j} = \{0,1\}^{N}$ be the  set of signals produced by cell-type $C_{j}$. For any gene $k$, $\SG_{j}(k) = 1 \iff  (C_{j}(k) = 1)\wedge(\SG(k)=1)$.

Parameter $\Padj\in [0,1]$ gives the probability of signal reception. We encode signal reception in a binary matrix $A _{2^N\times2^N}$. Cell-type $C_{i}$ receives all signals produced by cell-type $C_{j}$ if $A(j,i)=1$, where $A(j,i)=\Ber(\Padj)$. $C_{i}$ receives no signals from cell-type $C_{j}$ if $A(j,i)=0$ (Fig.\ref{fig:Fig1}(D)).
Cells can only receive signals from other cell-types present in the same time step. Let $T_{t} = \{0,1\}^{2^{N}}$ be a binary vector, where $T_{t}(i) =1 $ if cell-type $C_i$ is present in the time step $t$. $T_{t}$ represents the state of the organism at time step $t$.  
 For some cell-type $C_{i}$ present at time step $t$, let $\Csigi$ represent its state immediately after signal exchange. In cell-types that receive a signal, the corresponding genes are set to 1 (Fig.\ref{fig:Fig1}(F)). That is,
\begin{center}
$\Csigi(k) = 1$, if $(C_{i}(k)=1) \vee (\sum_{j=1}^{2^{N}}(A(j,i) \times \SG_{j}(k) \times T_{t}(j)) > 0)$
\end{center}

\subsubsection{Gene regulation}

A cell-type in the model need not be a fixed point (single cell-state) of the gene regulatory network, it can also be an oscillation (multiple cell-states) \citep{Xiadev169854}. In the latter case, the cell-type is represented by all cell-states that are part of the oscillation. We are only concerned with the set of cell-states in the stable state, and not with the sequence of cell-states in oscillations. 

Formally, a system with $N$ genes can have $n \leq 2^N$ stable cell-types \{$S_{1}, S_{2},...,S_{n}$\}; where $S_{x}$ is itself a collection of $n_x$ cell-states $\{C_{x_1}, ...C_{x_{n_x}}\}$ such that $x_1 < x_2 <...<x_{n_x}$. For any two cell-types $S_{x}$ and $S_{y}$, if $x<y$ then $x_1<y_1$. 

We encode gene regulation in a binary matrix  $GR _{2^N\times2^N}$. To generate $GR$ for a given organism, we pick the number of stable cell-types $n \leq 2^N$ according to uniform random distribution. First, we assign cell-states that form the basins of these stable cell-types: Cell-states are uniform randomly partitioned among the $n$ basins. We then choose cell-states that form the stable cell-type from within the corresponding basins. Let $B_x$ be a basin, then for some $j$ such that ($C_j \in B_x$), ($C_j \in S_x$) with probability 0.5. For all $i$ such that $C_{i} \in B_x$, $\GR(i,j) = 1$ if $(C_{j} \in S_x)$.

\subsubsection{Homeostatic organisms and their cell-type lineage graphs}
Let us consider an organism in state $T_{t}$ at time step $t$. Right after cell division, let the state of the organism be represented by $\Tdiv_t$. After division, the organism is composed of all the daughter cells produced in that time step. That is,
\begin{center}
$\Tdiv_{t}(i) = 1$, if $\exists j \leq 2^{N}$ s.t. $(T_{t}(j) = 1) \wedge (\CD(j,i) = 1)  $
\end{center}

These daughter cells exchange signals among themselves. Let $\Tsig_t$ represent the state of the organism right after signal exchange. Then,

\begin{center}
$\Tsig_{t}(i) = 1$, if $\exists j1 \leq 2^{N}$ s.t. $\Tdiv_{t}(j1) = 1$, where

    $\forall k$ s.t. $C_{i}(k) = 0$, $C_{j1}(k) = 0$, and 
    
    $\forall k$ s.t. $C_{i}(k) = 1$, $(C_{j1}(k) = 1)\vee(\sum^{2^N}_{j2=1}(A(j2,j1) \times \SG_{j2}(k) \times \Tdiv_t(j2))>0)$
\end{center}

The signals received by a cell-type activates its gene regulatory network. Gene regulation updates the set of cell-types according to the following expression: $\forall i \leq 2^{N},$
\begin{center}
$ T_{t+1}(i) = 1$, if $\exists j$ s.t. $(\Tsig_{t}(j)=1) \wedge (\GR(j,i)=1)$
\end{center}
Therefore, the organism is only composed of stable cell-types.
Let the system have $n \leq 2^{N}$ stable cell states. Then, we can equivalently represent the state of the organism at time step $t$ as a binary vector $T^{SC}_{t} = [0,1]^{n}$, such that for $x \in \{1,2,...n\}$.
\begin{center}
$T^{SC}_{t}(x) = 1 \iff (T_{t}(i) = 1) \wedge (\exists C_{i} \in S_{x})$
\end{center}

We call states of the organism such that $T^{SC}_{t+1} = T^{SC}_{t}$ homeostatic organisms (Fig.\ref{fig:Fig1}(F,G)).

We represent the homeostatic organism as a cell-type lineage graph. The nodes of the graph represent stable cell states that are present in the homeostatic organism, and directed edges represent lineage relationships between these stable cell states. Let the stable cell states $S_{x1}$ and $S_{x2}$ both be present in the final organism, and let them be represented by nodes $V_{a}$ and $V_{b}$ of the lineage graph respectively. Then, there is an edge from $V_{a}$ to $V_{b}$ if one of the daughter cells of $S_{x1}$ gives rise to $S_{x2}$ after one round of cell-signaling and gene regulation (Fig.\ref{fig:Fig1}(G)). That is,
\begin{center}
Let $C_{i} \in S_{x1}$ and $C_{l} \in S_{x2}$. 

Then, there is an edge $V_a \rightarrow V_b$ if

 $\exists j$ s.t. $\CD(i,j)=1$
 
and, in this organism, $\Csigj = C_{k}$

where $\GR(k,l) = 1$ 

\end{center}
\subsection{Assignment of topologies to lineage graphs}
We categorize lineage graphs into 6 topologies: unicellular, strongly connected component(SCC), cyclic, chain, tree and other directed acyclic graphs (DAG). We ignore self-edges while assigning these topologies. A lineage graph is called \textit{unicellular} if it has only a single node. For all other topologies, we used the networkx (version 2.2) module of Python3.6. A lineage graph is called \textit{SCC} if the graph has more than 1 node and contains a single strongly connected component, it is called \textit{cyclic} if the graph contains cycles and has more than one strongly connected component, it is called a \textit{chain} if networkx classifies it as a tree and the maximum in-degree and out-degree are 1, it is called a \textit{tree} if networkx classifies it as a tree and maximum in-dergee or out-degree is greater than 1, and it is called a \textit{DAG} if networkx classifies it as a directed acyclic graph but not a tree. 

\subsection{Lineage graph randomization protocol}
We represent a lineage graph with $e$ edges as a matrix $E_{e \times 2}$, where $E(i,1)$ and $E(i,2)$ represent the source and the target node of edge $i$ respectively. To randomize lineage graphs, we used a protocol that preserves in and out degrees of each node; we randomly choose pairs of edges from the graph and swap their target nodes. Let the randomized graph $\Erand$ be initially identical to $E$. Then,
\begin{center}
for any two edges of the lineage graph $i,j$, we propose a swap

$\Erand(i,2) = \Erand(j,2)$, and $\Erand(j,2) = \Erand(i,2)$

The swap is accepted if there is no edge $k$ such that

$(\Erand(k,1) = \Erand(i,1))\wedge(\Erand(k,2) = \Erand(j,2))$, or $(\Erand(k,1) = \Erand(j,1))\wedge(\Erand(k,2) = \Erand(i,2))$

\end{center}
The above condition ensures that the total number of unique edges in $E$ and $\Erand$ remain the same. We swap edges 1000 times for each lineage graph to randomize it.

\subsection{Independent and intrinsically independent cell-types}

We call a cell-type \textit{independent} if it has the same cell fate when grown outside the organism as it does when it is a part of the organism. The cell fate $\Cfate_i$ of some cell-type $C_i$ in the organism is given by the set of cell-types receiving an edge from the node $C_i$ in the organism's lineage graph. To decide whether a given cell-type $C_i$ is independent or not, we separate this cell-type from the rest of the organism, and allow it to undergo one round of cell division, signaling and gene regulation, according to the same matrices $\CD$, $\SG$, $A$ and $\GR$ that were used to generate the organism from which it was taken. Let us call the resulting set of cell-types $\Creg_i$. We call the cell $C_i$ independent if $\Creg_i$ is identical to $\Cfate_i$.

For some cell-types, the basis of their independence is an insensitivity to signals produced in the organism. In such a case, the set of signals produced by the daughter cells of the cell-type is sufficient to satisfy the maximum set of signals that each of the daughter cells can receive.

Let the set of daughter cells of cell-type $C_i$ in an organism be $D_i$. $\forall C_j \in D_i$ let $\Recall_j$ represent the maximal set of signals that it can receive, when all $2^N$ possible cell-types are present together. i.e., For all signaling molecules k such that $\SG(k) =1$,
\begin{center}
    $\Recall_j(k) = 1$, if $\Sigma^{2^N}_{l=1}(A(l,j) \wedge (C_l(k) = 1))$
\end{center}

And, let $\Rec^D_j$ be the set of signals it receives from within the set of cells $D_i$. i.e., 
\begin{center}
    $\Rec^D_j(k) = 1$, if $\Sigma_{C_l \in D_i} (A(l,j) \wedge (C_l(k) = 1))$
\end{center}

If, for all $C_j \in D_i$, $\Recall_j$ = $\Rec^D_j$,
$C_i$ is \textit{intrinsically independent}.

\subsection*{Acknowledgements}
We thank Luca Peliti, Albert Libchaber and Mukund Thattai for useful discussions, and John McBride for assistance with writing Python code for analysis. This work was supported by the taxpayers of South Korea through the Institute for Basic Science, Project Code IBS-R020-D1.

\subsection*{Author Contributions}
S.M. conceived the project, developed code for simulations and performed analysis; S.M. and T.T. designed research; S.M. and T.T. wrote the paper.

\subsection*{Declaration of interests}
The authors declare no competing interests.

\vspace{11mm}
\bibliographystyle{unsrtnat} 
\bibliography{references}


\end{multicols}

\vspace{20mm}
\section{Supplementary material}
\renewcommand{\thefigure}{S\arabic{figure}}
\setcounter{figure}{0}
\subsection{Dissection of the effect of parameters on graph size} 
\begin{figure}[htbp!]
    \centering
    \includegraphics[scale=0.5]{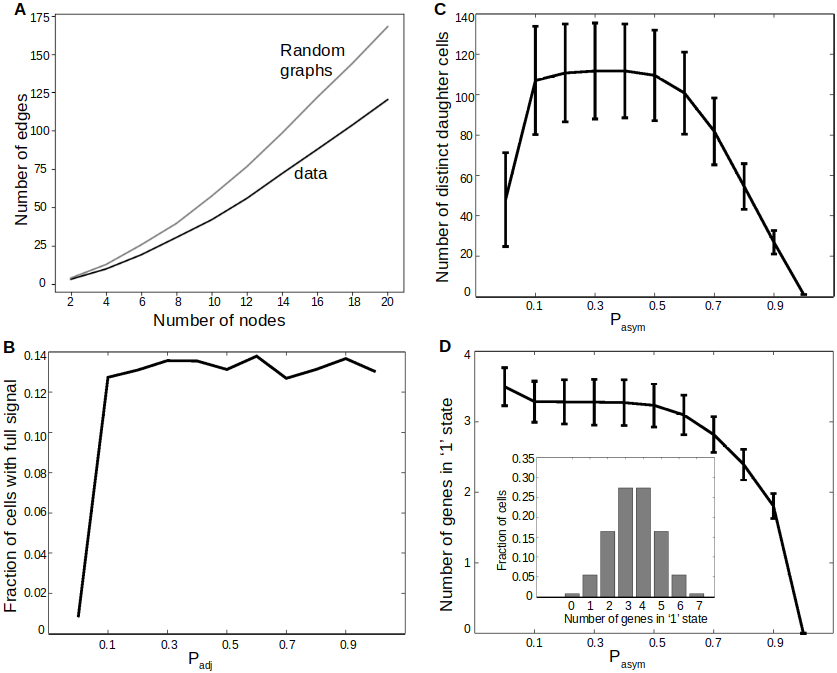}
    \caption{\textbf{Effect of parameters on graph size.} \textbf{(A)} A comparison of growth rate of the number of edges with number of nodes in lineage graphs of our data, versus that expected of Erdos-Renyi random graphs. \textbf{(B)} Effect of $\Padj$ on signal reception. At each value of $\Padj$, 1000 random signaling vectors $\SG$ for $N=7$ organisms, generated at $\Psig = 0.5$ were used. In each organism, the set of signals received by a randomly chosen cell-type, from all $2^N$ possible cell-types in the system was assessed. The horizontal axis represents $\Padj$, and the vertical axis represents the fraction of cell-types out of 1000, that received all possible signals. \textbf{(C,D)} Effect of $\Pasym$ on number of nodes in lineage graphs. At each value of $\Pasym$, 10,000 'organisms' with $N=7$ genes, composed of randomly chosen cell-types were used to generate these graphs.\textbf{(C)} Average number of distinct daughter cells produced in an organism as a function of $\Pasym$. Error-bars indicate standard deviation. \textbf{(D)} Average number of genes in '1' state in daughter cells as a function of $\Pasym$. Error-bars indicate standard deviation. Inset: frequency of cell-types with $N=7$ genomes with different numbers of genes in a '1' state (horizontal axis).}
    \label{fig:supp_fig1}
\end{figure}

We compared the properties of lineage graphs in our data with those generated with Erdos-Renyi random graphs (ER graphs) of similar size. ER graphs are generated using a fixed probability, $p$ of any two nodes in the graph being connected by an edge \citep{erdos59a}. Therefore, on average, the number of edges in a graph with $n$ nodes is proportional to $n^2$. Increasing the number of nodes to $c*n$ increases the number of edges to $c^2 * n^2$. We determined the number of edges in lineage graphs with $n = [1,2,3,4,5,6,7,8,9,10]$ nodes, and calculated the number of edges expected in ER graphs with $n = [2,4,6,8,10,12,14,16,18,20]$ nodes. Compared to ER graphs, the rate of growth of number of edges in lineage graphs in our data is noticeably slower (Fig.\ref{fig:supp_fig1}(A)).

The number of nodes in lineage graphs decreases sharply with the parameter $\Padj$ (Fig.\ref{fig:Fig2}(E)). We show here that this occurs because even at low values of $\Padj$, cell-types in organisms are connected enough that the fraction of cell-types receiving all signals produced in the organisms reaches a maximum (Fig.\ref{fig:supp_fig1}(B)).

The effect of the parameter $\Pasym$ on the number of nodes in lineage graphs can be explained in terms of its effect on the number of distinct daughter cell-types produced (Fig.\ref{fig:supp_fig1}(C)). The number of distinct daughter cells produced at different values of $\Pasym$ is related to the average fraction of genes in a 1 state in these daughter cells. Among all possible cell-types with $N$ genes, most cell-types tend to have about half their genes in a 1 state, and very few cell-types contain fewer, or more genes in a 1 state (Fig.\ref{fig:supp_fig1}(D:inset)). Therefore, when $0.2 < \Pasym < 0.6$, where on average, daughter cells have about half their genes in a 1 state, organisms produce the most number of distinct daughter cells, and at $\Pasym < 0.2$ and $\Pasym > 0.6$, fewer distinct daughter cells are produced (Fig.\ref{fig:supp_fig1}(D)).

\subsection{Randomization of lineage graphs}
\begin{figure}[htbp!]
    \centering
    \includegraphics[scale=0.38]{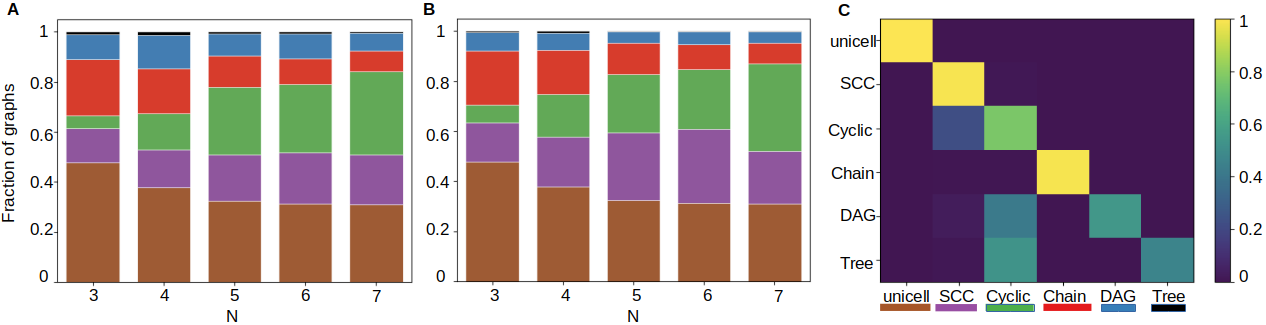}
    \caption{\textbf{Distribution of topologies of randomized graphs.} The data used here is smaller than, but overlapping with, that used in the main paper. 2373473 graphs were used here. \textbf{(A,B)} Stacked histograms of graph topologies. Different topologies are represented by different colours: unicellular: brown, SCC: purple, cyclic: green, chain: red, DAG: blue and tree: black. Heights of coloured blocks indicate the proportion of graphs of the corresponding topology. \textbf{(A)} lineage graphs generated by the model, \textbf{(B)} randomized lineage graphs. \textbf{(C)} 2-D histogram representing conversions of  graph topology due to randomization. Rows indicate the topologies of original graph and columns indicate the topologies of randomized versions. Intensity of colours in the histogram indicates the fraction of conversions of each type, according to the colourbar given alongside.}
    \label{fig:figs2}
\end{figure}

We randomized lineage graphs generated with our model while keeping node in-degrees and out-degrees unchanged. Topology distribution largely remains unchanged upon randomization (Fig.\ref{fig:figs2}(A,B)), and not many graphs change their topology upon randomization (Fig.\ref{fig:figs2}(C)). Although, we find that the proportion of acyclic graphs decreases slightly, from 24\% in model generated graphs, to 19\% in randomized graphs.

\subsection{Characteristics of lineage graphs with different topologies: graph size}
\begin{figure}[htbp!]
    \centering
    \includegraphics[scale=0.47]{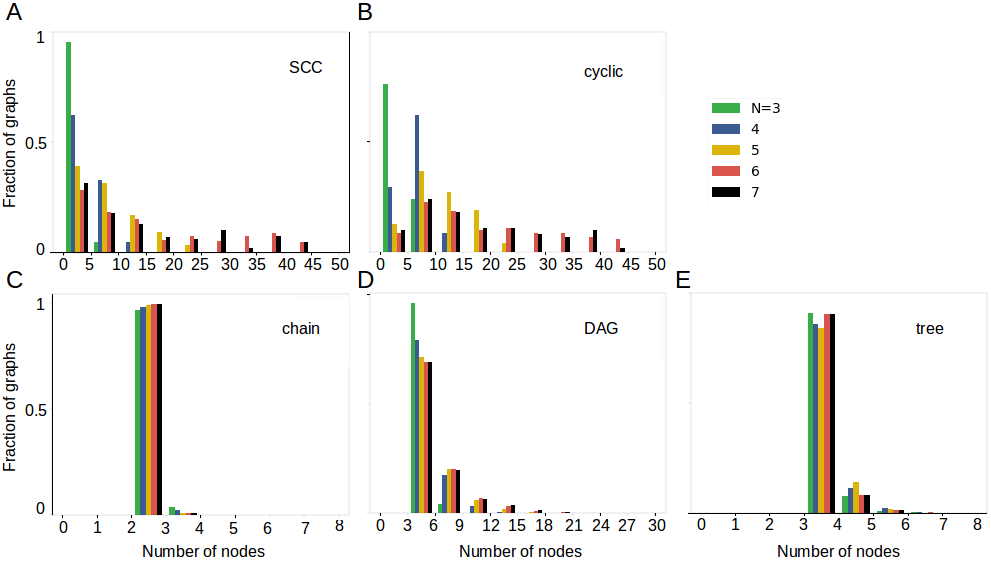}
    \caption{\textbf{Graph size distributions for different topologies.} The horizontal axis indicates number of nodes in lineage graphs and the vertical axis indicates the normalized frequency of graphs. Histogram bin sizes are as follows: \textbf{(A,B)} 5, \textbf{(C)} 1, \textbf{(D)} 3, \textbf{(E)} 1.}
    \label{fig:figs3}
\end{figure}

Different graph topologies are different in their graph size distributions. While SCC and cyclic graphs span a large range of graph sizes (Fig.\ref{fig:figs3}(A,B)), trees and chains tend to be notably small (Fig.\ref{fig:figs3}(C,E)). DAG type graphs can have moderately large number of nodes (Fig.\ref{fig:figs3}(D)). 

\subsection{Characteristics of tree-type lineage graphs}
Tree-type graphs can be further characterized as divergent or convergent trees. We call graph nodes with in-degrees > 1 convergent, and nodes with out-degrees > 1 divergent. Note that by this definition, the same node is allowed to be both convergent and divergent. For some tree-like graph with $n$ nodes and $n_e$ edges, let $in_i$ and $out_i$ be the in-degree and out-degree of the $i^{th}$ node respectively. We define for this graph a number $g_c$ as the sum of in-degrees of all convergent nodes, and a number $g_d$ as the sum of out-degrees of all divergent nodes, i.e.;

\begin{center}
     $g_c = \Sigma in_i$, $\forall i$ s.t. $in_i > 1$,
     
     $g_d = \Sigma out_i$, $\forall i$ s.t. $out_i > 1$
\end{center} 

\begin{figure}[htbp!]
    \centering
    \includegraphics[scale=0.36]{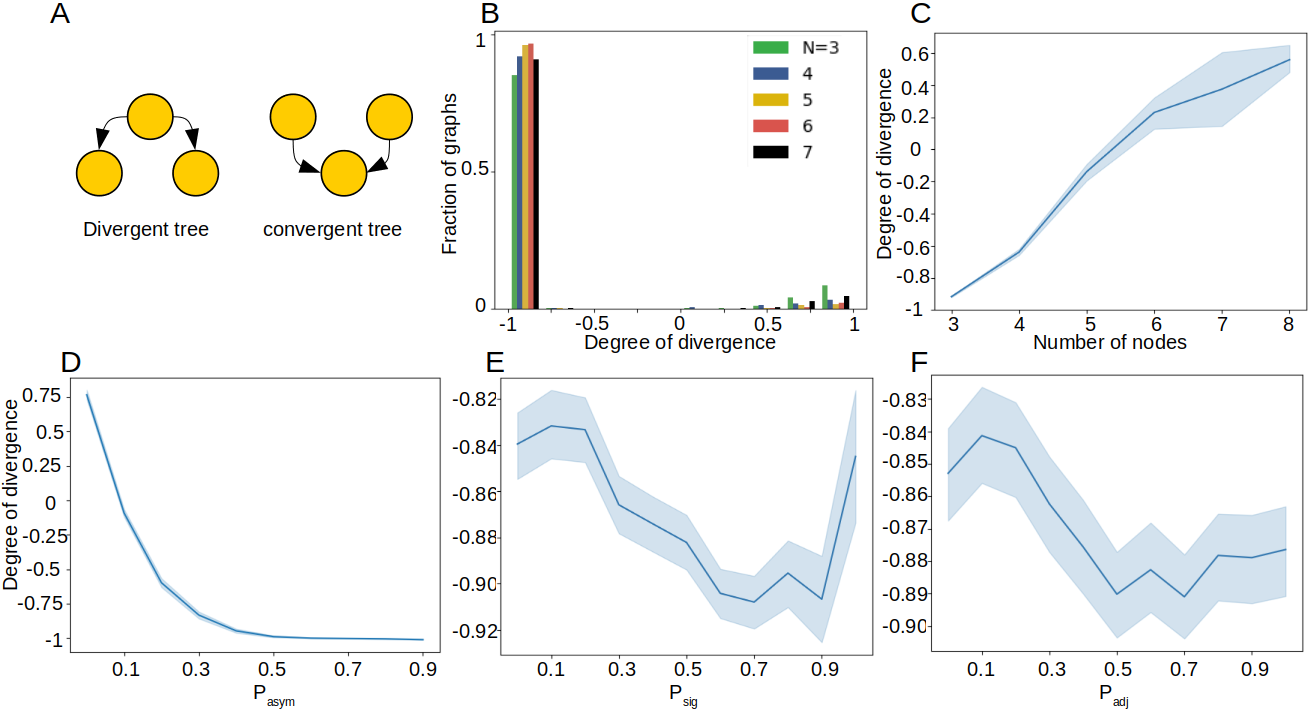}
    \caption{\textbf{Properties of tree-type graphs.} \textbf{(A)} Schematics of a divergent tree like lineage graph and a convergent tree like lineage graph. Yellow circles represent cell-types and edges represent lineage relationships. \textbf{(B)} Histogram of degrees of divergence for tree-like graphs in our data with different $N$. \textbf{(C,D,E,F)} Average degree of divergence in our data as a function of \textbf{(C)} number of nodes in lineage graphs, \textbf{(D)} $\Pasym$, \textbf{(E)} $\Psig$, \textbf{(F)} $\Padj$. Shaded regions indicate standard deviation.}
    \label{fig:figs4}
\end{figure}

We define the degree of divergence of this graph as $(g_d - g_c)/n_e$. For a perfectly divergent tree, such as the tree to the left in Fig.\ref{fig:figs4}(A), the degree of divergence is 1. And for a perfectly convergent tree (e.g. the tree to the right in Fig.\ref{fig:figs4}(A)), degree of convergence is -1. We find that most tree-like graphs in our data tend to be more convergent than divergent (Fig.\ref{fig:figs4}(B)). Lineage graphs that are divergent lead to an increase in cell-type diversity starting from a few initial cell-types. Lineage graphs of real organisms are believed to be divergent trees. Larger trees tend to be more divergent (Fig.\ref{fig:figs4}(C)). Degree of divergence decreases as $\Pasym$ increases, it is relatively insensitive to $\Psig$ and $\Padj$.

\subsection{Characteristics of DAG-type lineage graphs}
\begin{figure}[htbp!]
    \centering
    \includegraphics[scale=0.45]{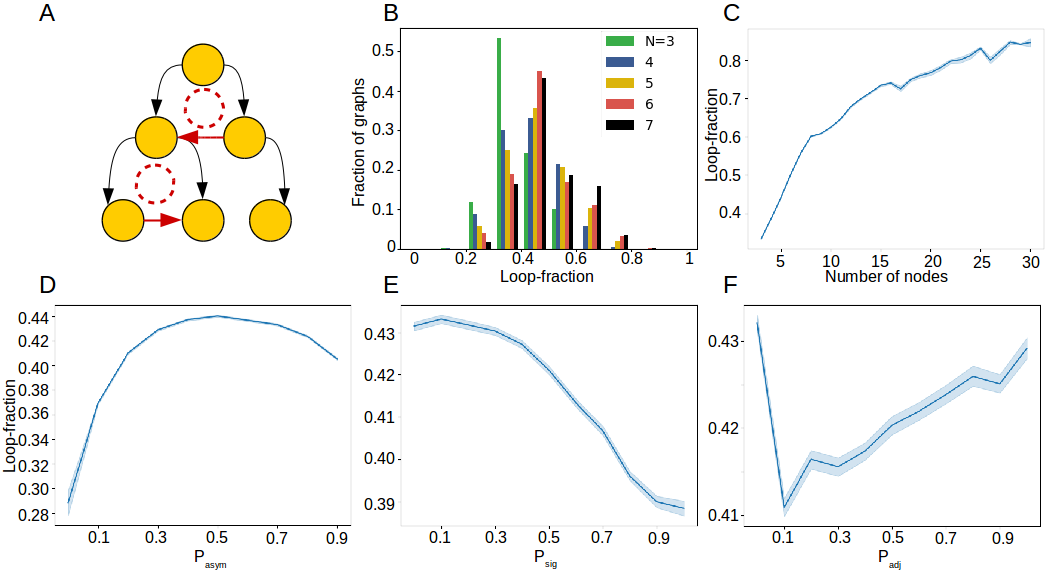}
    \caption{\textbf{Properties of DAG-type graphs.} \textbf{(A)} Schematic of DAGs. Yellow circles represent cell-types, and edges represent lineage relationships. Red edges forms loops in the DAG.  \textbf{(B)} Histogram of loop-fraction of DAGs in our data. Loop-fraction is defined as the fraction of edges in a DAG that form loops. Histogram bins are of size 0.1. \textbf{(C,D,E,F)} Average loop-fraction of DAG type graphs in our data as a function of \textbf{(C)} number of nodes in lineage graphs, \textbf{(D)} $\Pasym$, \textbf{(E)} $\Psig$, \textbf{(F)} $\Padj$. Shaded regions indicate standard deviation.}
    \label{fig:figs5}
\end{figure}

DAG-type graphs differ from tree-like graphs in having edges that link different branches. If the edges in the DAG are rendered undirected, these edges are parts of cycles, or loops (Fig.\ref{fig:figs5}(A)). The number of such edges in DAGs can be determined by subtracting the number of edges in the spanning tree of the graph from the total number of edges. For a graph with $n$ nodes, the spanning tree has $n-1$ edges. For a given DAG-type graph, we call the fraction of its edges that forms loops, its loop-fraction. Loop-fractions of DAG-type lineage graphs indicate the level of trans-differentiation. DAG-type graphs in our data have high loop-fractions (Fig.\ref{fig:figs5}(B)), and loop-fraction increases with graph size (Fig.\ref{fig:figs5}(C)). 

\subsection{Distribution of regenerative capacities}
\begin{figure}[htbp!]
    \centering
    \includegraphics[scale=0.5]{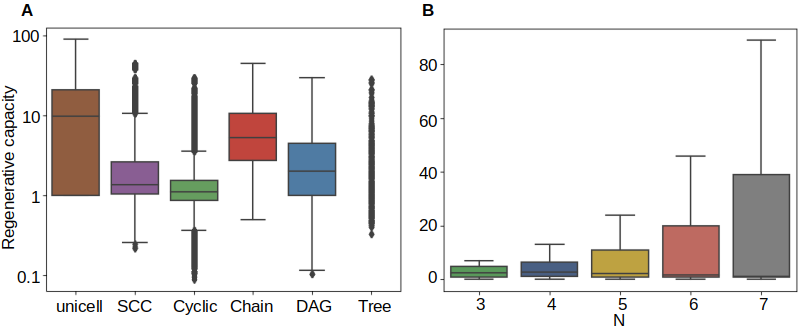}
    \caption{\textbf{Box plots for regenerative capacity of lineage graphs} \textbf{(A)} across different topologies, \textbf{(B)} across number of genes $N$. Boxes represent quartiles of the data set. Lines inside the box shows the median, while whiskers show the rest of the distribution. Outliers are shown as diamonds. Most tree-like lineage graphs have a regenerative capacity of 0, therefore the box for these graphs is not visible.}
    \label{fig:figs6}
\end{figure}

In order to infer whether a lineage graph is regenerative, we only look at whether its regenerative capacity is greater than 1, or not. In Fig.\ref{fig:figs6}(A), we show the spread of regenerative capacities for different topologies. For most topologies, median regenerative capacity is greater than 1. The actual value of regenerative capacity is less meaningful, except in the case of tree-type graphs, where most trees have a regenerative capacity of 0. This implies that most trees contain no pluripotent cells.
We also find that while median regenerative capacity decreases with $N$, the range of regenerative capacities increases with $N$ (Fig.\ref{fig:figs6}(B)).

\subsection{Organisms in the model have short regeneration trajectories}
\begin{figure}[htbp!]
    \centering
    \includegraphics[scale=0.5]{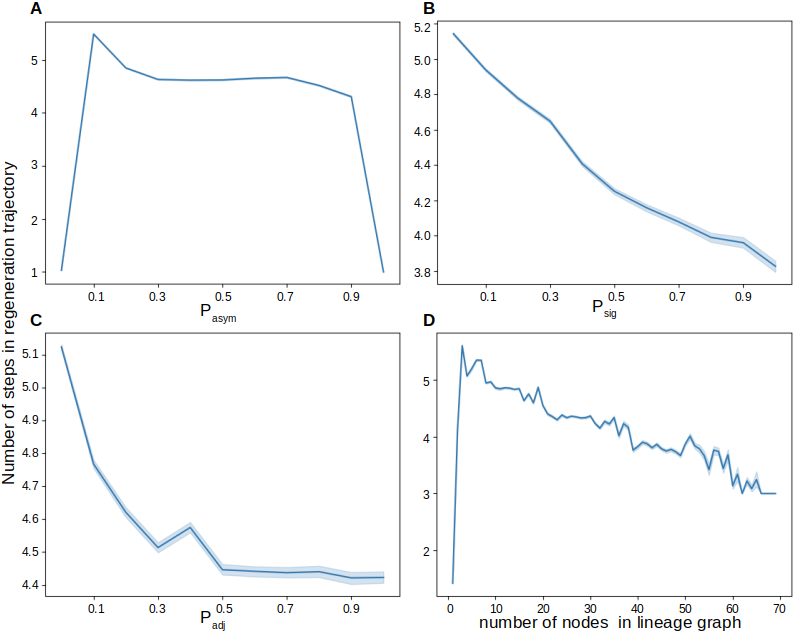}
    \caption{\textbf{Average regeneration trajectory lengths} \textbf{(A)} as a function of $\Pasym$, \textbf{(B)} as a function of $\Psig$, \textbf{(C)} as a function of $\Padj$, \textbf{(D)} as a function of number of nodes in lineage graph. Shaded regions represent standard deviation.}
    \label{fig:figs7}
\end{figure}

Regeneration trajectories from pluripotent cells to the homeostatic organism tend to be short (Fig.\ref{fig:figs7}). On average, trajectory lengths do not depend on whether the pluripotent cell is independent or not. At first, the observation that trajectory lengths decrease with graph size (Fig.\ref{fig:figs7}(D)), might seem incongruous. But it can be understood intuitively by seeing that number of nodes in lineage graphs is proportional to the number of distinct daughter cells produced in every step of development. Since the same rules are followed for development and regeneration in the model, we can expect the the number of new cell-types added at each step of the regeneration trajectory is larger for organisms with more nodes in their lineage graphs.  

\subsection{Intrinsically independent cell-types are enriched in lineage graphs}
\begin{figure}[htbp!]
    \centering
    \includegraphics[scale=0.5]{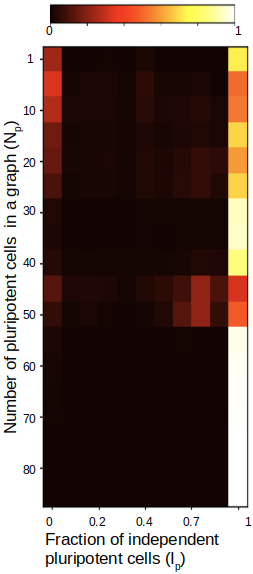}
    \caption{\textbf{Independent pluripotent cell-types} 2-D histogram indicating the fraction of independent pluripotent cell-types in homeostatic organisms. Intensity of colours in the histogram indicate the fraction of organisms with $N_p$ pluripotent cell-types, $I_p$ of which are independent, according to the colourbar given on top.}
    \label{fig:figs8}
\end{figure}

\begin{figure}[htbp!]
    \centering
    \includegraphics[scale=0.5]{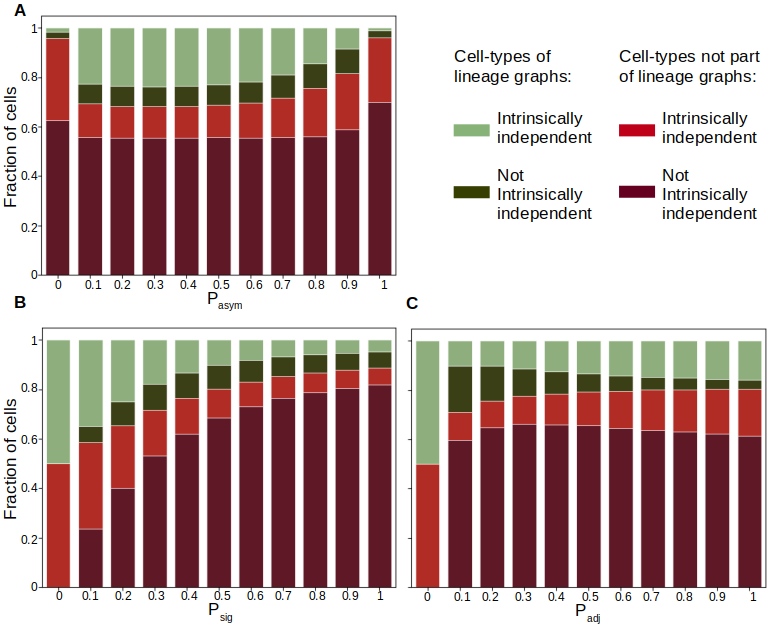}
    \caption{\textbf{Stacked histograms showing intrinsic independence of cell-types} \textbf{(A)} as a function of $\Pasym$, \textbf{(B)} as a function of $\Psig$, \textbf{(C)} as a function of $\Padj$. Different cell-type categories are represented with different colours. Cell-types not part of organisms are represented in reds; intrinsically independent: bright red, not intrinsically independent: dark red. Cell-types found in organisms are represented in greens; intrinsically independent: light green, not intrinsically independent: dark green. Heights of colored blocks represent the proportions of corresponding cell-types.}
    \label{fig:figs9}
\end{figure}

We find in the model that the cell-fate of most pluripotent cells is independent of cellular context (Fig.\ref{fig:figs8}).
We wondered whether the large number of independent cell-types in lineage graphs in our data could be attributed to an insensitivity of these cell-types to signals from other cell-types. Alternatively, these cell-types could be independent despite being responsive to signals from other cell-types. We find that the former case tends to be true. We call a cell-type \textit{intrinsically independent} if the full set of signals that can potentially be received by each of its daughter cells is already satisfied by signaling among these daughter cells themselves. In other words, no further external signals can influence the fates of the daughter cells of intrinsically independent cell-types. We calculated the fraction of intrinsically independent cell types across all $2^N$ possible cell-types across all systems in our data. We find that cell-types that are part of lineage graphs are much more likely to be intrinsically independent irrespective of parameter region (Fig.\ref{fig:figs9}). Thus cell-types in lineage graphs are predisposed to be independent.But, not all independent cell-types in lineage graphs are intrinsically independent (overall, about 20\% the independent cells across all lineage graphs are not intrinsically independent), and not all independent cell-types are pluripotent (Fig.\ref{fig:Fig4}(F)). 

\subsection{Drosophila segement polarity network expressed in terms of the generative model}
In Drosophila embryos, segment polarity genes maintain borders of parasegments, which are 4 cells wide. Within each parasegment, the polarity genes are expressed in characteristic stripes. In \cite{albert2003topology}, the authors demonstrated that the gene regulatory network responsible for the pattern of gene expression in this system can be modeled as a Boolean logical network. In the following, we examine the Drosophila segment polarity network in terms of our generative model. 

\begin{figure}[htbp!]
    \centering
    \includegraphics[scale=0.5]{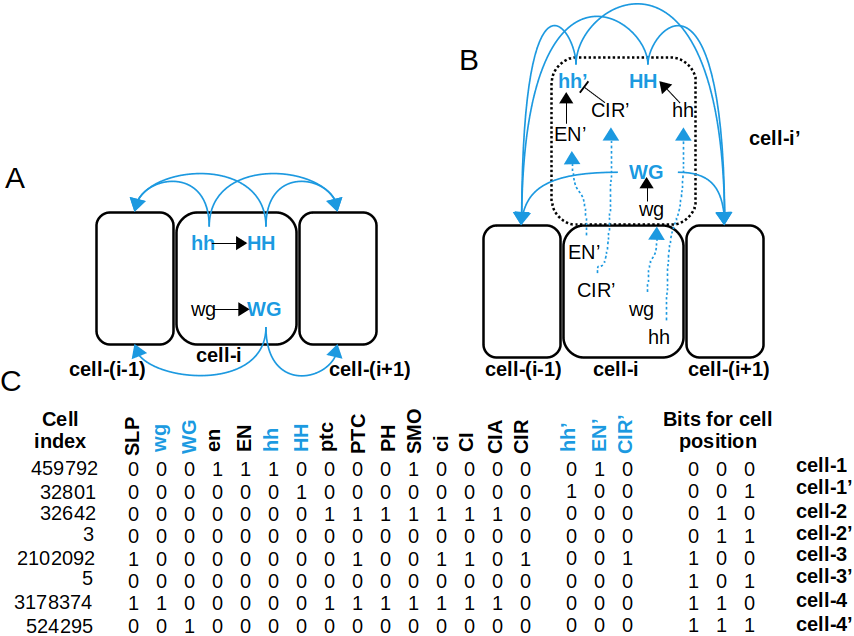}
    \caption{\textbf{Modified segment polarity network} \textbf{(A)} Signaling in the original model, as in \cite{albert2003topology}). Signaling molecules are labeled in blue. Blue edges represent signal transduction, and black edges represent 'gene'-regulation. \textbf{(B)} Modified structure of segment polarity network. We introduce a new cell, shown here with a dotted outline, adjacent to the original cell, which acts like an insulated compartment of this cell. All signals are transmitted via this new cell to neighbouring cells. \textbf{(C)} Steady state of the modified network that corresponds to wild-type stripe pattern in Drosophila, as reported in \cite{albert2003topology}. Each row is a cell-type, and columns represent states of ‘genes’. 1 implies presence of the gene product, and 0 implies absence of the gene product. There are 21 genes in the system: the first 15 genes are the original mRNAs and proteins used to construct the regulatory network in \cite{albert2003topology}, and the next 3 genes represent ‘mirrors’ of hh, EN and CIR which are used for signaling purposes. The last 3 ‘genes’ encode the position of the cell along the anterio-posterior axis; cell-1 is the most anterior and cell-4 is the most posterior. Cells 1'-4' represent the new cells we introduce for signal transduction. Each cell is indexed by the decimal number obtained upon converting the corresponding 21-length binary vector. }
    \label{fig:figs10}
\end{figure}

The network consists of 15 nodes: en, wg, hh, ptc and ci represent mRNAs, and SLP, EN, WG, HH, PTC, SMO, PH, CI, CIA and CIR represent proteins. Of these, WG, hh and HH act as signals. Signaling molecules HH and WG do not participate in regulation within the cells that produce them, rather they act only in cells that receive them as signals. In order to incorporate this feature, we represent each cell in the parasegment as two model cells; production of all non-signal molecules takes place in one of the cells, and molecules responsible for regulation of signal molecule production are exported to the second cell, from which signal molecules are secreted (Fig.\ref{fig:figs10}(A,B)). In this sense, the second cell acts as a special compartment which insulates the gene network in the first cell from regulation by signal molecules produced within the same cell.

In this system, signals are exchanged only between neighbouring cells. Accordingly, in our model, cell positions can be expressed as additional 'genes', whose states do not change. For example, to express the positions of the $4 \times 2$ cells in this system, we use 3 additional 'genes' (Fig.\ref{fig:figs10}(C)). In this system, signal exchange only depends on these 'positional genes', and does not depend on the states of the other genes. 

\begin{figure}[htbp!]
    \centering
    \includegraphics[scale=0.5]{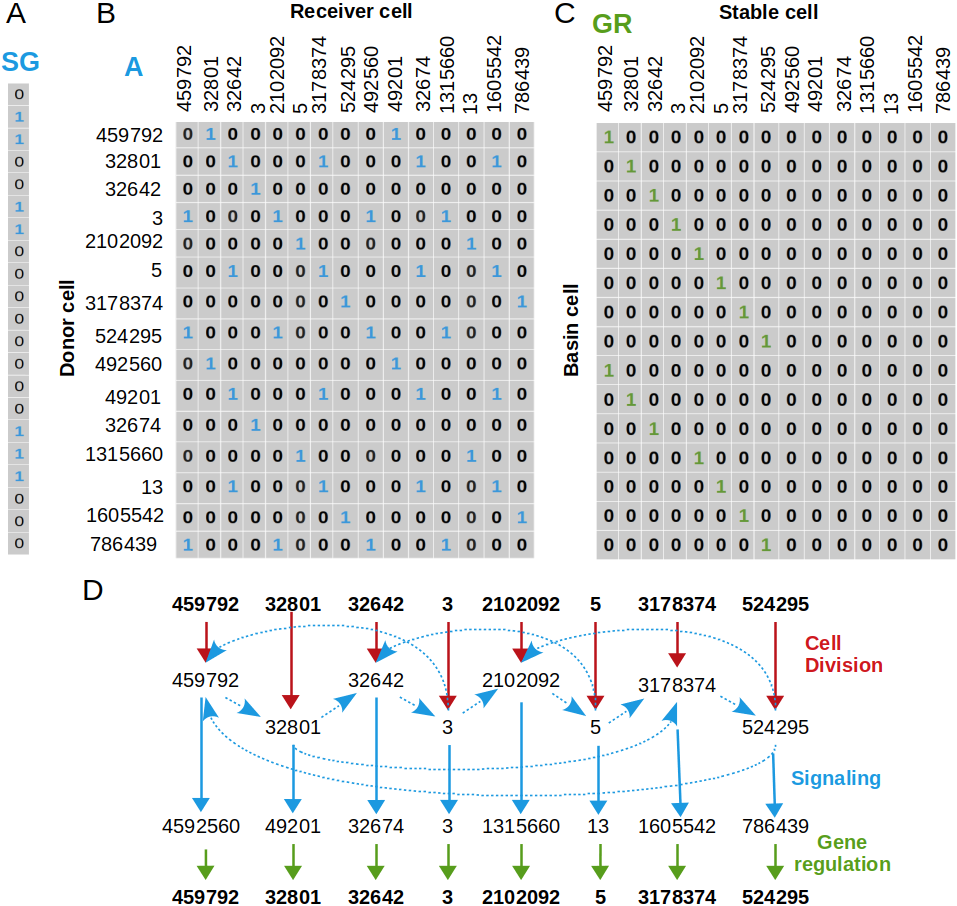}
    \caption{\textbf{Developmental rules matrices for the modified segment polarity network} \textbf{(A)} signaling vector $\SG$, \textbf{(B)} Cellular adjacency matrix $A$. Note that cellular adjacency is completely determined by cell positions; Cells-1-4 only pass on molecules to cells-1'-4' respectively, and a cell-i' only passes on signals to cell-(i-1) and cell-(i+1). Periodic boundary conditions are employed here, which implies that cell-1 and cell-4 are neighbours. \textbf{(C)} Gene regulation matrix $GR$. The first 8 cell-types correspond to the stable state, as in Fig.\ref{fig:figs10}(C). In \textbf{B} and \textbf{C}, only the relevant parts of the rules matrices are shown. The full matrices are of size $2^{21} \times 2^{21}$. \textbf{(D)} Schematic diagram of signaling and gene regulation in determining the wild-type steady state of the \textit{Drosophila} segment polarity network. Numbers represent the indices of different cell-types. Red arrows represent cell-division, which is symmetric in this case. Dashed blue arrows represent signal exchange among cell-types and solid blue arrows represent changes in cell-types due to signal exchange. Green arrows represent gene regulation.}
    \label{fig:figs11}
\end{figure}

Thus, our model is capable of expressing spatial arrangement of cells, and complex signaling mechanisms, although, it comes at the cost of an increase in system size. In Fig \ref{fig:figs11}, we show the signaling vector $\SG$, and portions of the cellular adjacency  matrix $A$, and gene regulation matrix $GR$ relevant to the steady state of the wild-type segment polarity network. The authors assume symmetric cell-division in \cite{albert2003topology}, and we do the same; therefore we do not show the cell-division matrix $\CD$ here.

\end{document}